# Querying Fault and Attack Trees: Property Specification on a Water Network

Stefano M. Nicoletti, University of Twente

Milan Lopuhaä-Zwakenberg, University of Twente

E. Moritz Hahn, University of Twente

Mariëlle Stoelinga, University of Twente and Radboud University

*SUMMARY & CONCLUSIONS*

We provide an overview of three different query languages whose objective is to specify properties on the highly popular formalisms of fault trees (FTs) and attack trees (ATs). These are *BFL*, a Boolean Logic for FTs, *PFL*, a probabilistic extension of *BFL* and *ATM*, a logic for security metrics on ATs. We validate the framework composed by these three logics by applying them to the case study of a water distribution network. We extend the FT for this network – found in the literature – and we propose to model the system under analysis with the *Fault Trees/Attack Trees* (FT/ATs) formalism, combining both FTs and ATs in a unique model. Furthermore, we propose a novel combination of the showcased logics to account for queries that jointly consider both the FT and the AT of the model, integrating influences of attacks on failure probabilities of different components. Finally, we extend the domain specific language for PFL with novel constructs to capture the interplay between metrics of attacks – e.g., "cost", success probabilities – and failure probabilities in the system.

## 1 INTRODUCTION

Critical infrastructure systems must operate safely and securely. Fault tree analysis (FTA) [1,2] is a widespread method used for risk assessment of these systems. Developed in the early '60s [3], fault trees (FT) are directed acyclic graphs (DAGs) that model how component failures arise and propagate through the modelled system, eventually leading to system level failures. Leaves in a FT represent *basic events* (BEs), i.e. elements of the tree that need not be further refined. Once these fail, the failure is propagated through the *intermediate events* (IEs) via *gates*, to eventually reach the *top level event* (TLE), which symbolizes system failure. In FTA, typically one identifies the *minimal cut sets* (MCSs) of a FT, i.e. minimal sets of BEs that, when failed, cause the system to fail. One can also identify *minimal path sets* (MPSs), i.e. minimal sets of BEs that – when operational – guarantee that the system will remain operational. FTs are a required analysis methodology by, e.g., the Federal Aviation Administration, the Nuclear Regulatory Commission, the ISO 26262 standard [4] for autonomous driving and for software development in aerospace systems.

Attack trees (ATs) [5] are the security counterpart of FTs: hierarchical diagrams that offer a flexible modelling language to assess how systems can be attacked. As for FTs, ATs are widely employed both in industry and academia: they are part of many system engineering frameworks, e.g. *UMLsec* [6] and *SysMLsec* [7, 8], and are supported by industrial tools such as Isograph's *AttackTree* [9].

### 1.1 Combining Fault and Attack Trees

Due to their popularity, numerous combinations and extensions of FTs and ATs have been proposed. Recent surveys [10, 11] attest that at least seven such combinations/extensions are popular in the literature: *Extended Fault Trees* or *Fault Trees/Attack Trees* (FT/ATs) [12], *Component Fault Trees* (CFTs) [13], *Attack-Fault Trees* (AFTs) [14], *State/Event Fault Trees* (SEFTs) [15], *Failure-Attack-CounTermeasure* (FACT) *Graphs* [16], *Boolean Driven Markov Processes* (BDMPs) [17] and *Attack Tree Bow-ties* (ATBTs) [18].

In this paper, we focus our attention on FT/ATs. These model the intuition that malicious actors often try to induce a failure of some components in a system, in order to render it non-operational: in doing so, they offer a sensible way of combining FTs and ATs. FT/ATs model these situations by replacing one or multiple BEs in the FT with the root of an AT, symbolizing paths that an attacker can take to cause failure in one or more basic components in an FT.

### 1.2 Querying Fault and Attack Trees

Despite their popularity, however, little work has been done on developing tailored languages that enable practitioners to specify flexible properties on FTs and ATs. Only very recent work addressed this issue, by proposing three different logics tailored to FTs and ATs, accompanied by model checking algorithms that can check the truth value of formulae.

*Boolean Fault tree Logic.* Our previous work [19] proposed a *Boolean Fault tree Logic* (BFL) with which practitioners can: 1. set evidence to analyze what-if scenarios, e.g., what are the MCSs, given that BE *A* or subsystem *B* has failed? What are the MPSs given that *A* or *B* have not failed? 2.

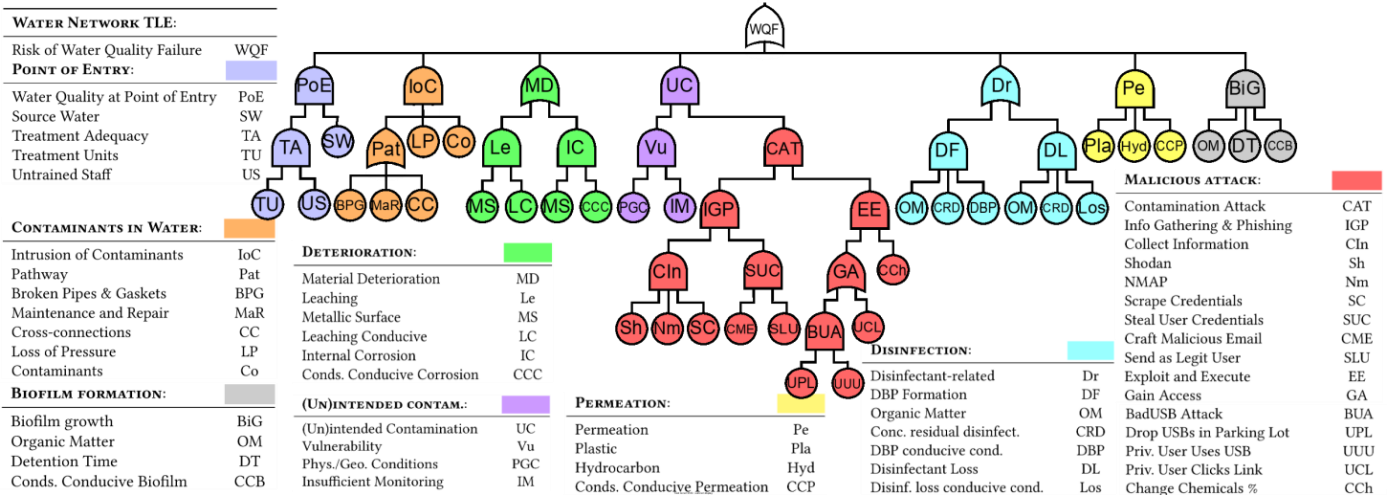

*Figure 1 – Fault tree with attacks (attack tree in red) for a water distribution network. Intermediate events are inside gates.*

check whether two FT elements are independent or if they share a child that can influence their status. 3. check whether the failure of one (or more) element *E* always leads to the failure of the TLE. 4. set upper/lower boundaries for failed elements, e.g., would element *E* always fail if at most/at least two out of *A*, *B* and *C* were to fail? Moreover, if a property does not hold, the BFL framework generates counterexamples, to show why the property fails.

*Probabilistic Fault tree Logic.* Extending the previous framework, [20] presents a *Probabilistic Fault tree Logic* to further enhance quantitative analysis capabilities, as probabilities are the prime quantitative metric on FTs. With PFL, one: 1. can check whether the probability of a given element (potentially conditioned by another one) respects a certain threshold, 2. can set the value of one BE in complex formulae to an arbitrary probability value, 3. can check if two BEs/IEs are stochastically independent, 4. can also return probability values for given formulae, possibly mapping single elements to an arbitrary probability value. Furthermore, [20] presents *LangPFL*, a domain specific language for PFL that propels the usability of this framework, allowing easier property specification on FTs.

*A Logic for Attack Tree Metrics.* Concerning ATs, [21] develops a *Logic for Attack Tree Metrics* (ATM) to specify a variety of quantitative security properties on these models: the authors present a general framework that considers *security metrics*, such as "cost" of an attack, "probability" of getting attacked and "skill" of a malicious actor. With ATM, one: 1. can reason about successful/unsuccessful attacks; 2. can check whether metrics, such as the cost, are bounded by a given value on single attacks; 3. can compute metrics for a class of attacks and 4. perform quantification over all possible attacks. Note that because ATM uses a general algebraic framework, it allows for the analysis of many different metrics [22].

### 1.3 Our Contribution

In this paper we propose an extended version of the FT in [23] that models a water distribution network. We enrich this model by providing a FT/AT showcasing a malicious attack that intends to contaminate water in the network. This scenario is not unlikely, as testified by recent news that see a Florida water treatment facility hacked using a dormant remote access software [24]. Furthermore, to validate the framework composed by BFL, PFL and ATM we showcase property specification for the FT and the AT composing the model. Moreover, we propose a novel combination of these logics and present joint property specification for the FT/AT model. Finally, we extend LangPFL – the domain specific language for PFL presented in [20] – to support different metrics and the specification of queries on ATs and FT/ATs.

### 2 CASE STUDY: WATER DISTRIBUTION NETWORK

The case study we are analyzing considers a water distribution network that might be subject to a contamination attack. The FT/AT in *Figure 1* represents a water distribution network: the TLE for the FT/AT models the risk of *Water Quality Failure* (WQF), that is refined via an OR-gate. Children of this gate are further refined in different subtrees. From left to right, we find an indigo AND-gate representing *Water Quality at Point of Entry* (PoE), an orange AND-gate for the *Intrusion of Contaminants* (IoC), a green OR-gate refining *Material Deterioration* (MD), a violet AND-gate for *(Un)intended Contamination* (UC), a light blue OR-gate representing *Disinfectant-related* (DR) risks, a yellow AND-gate for *Permeation* (Pe) risks and a grey AND-gate, refining events related to *Biofilm growth* (BiG).

Subtrees from the original FT are extended with an AT represented by an AND-gate for a malicious *Contamination attack* (CAT), in red. This AT refines one of the BEs present in the original FT model from [23] that generically represented a *Threat*. In our model, for the contamination attack to be successful, a malicious actor must perform *Information Gathering & Phishing* (IGP) to *Collect Information* (CIn) on the target infrastructure and to *Steal User Credentials* (SUC). Furthermore, the attacker must hit the target with an *Exploit, and Execute* (EE) the attack by *Changing Chemicals* (CCh) percentages in the water. To successfully execute this plan, the malicious actor can *Gain Access* (GA) by letting a *Privileged*

*User Click the Link* (UCL) of his/her malicious email, or by successfully executing a *BadUSB Attack* (BUA). Note that BEs shared between multiple IEs have a dashed border in *Figure 1*.

## 3 QUERYING FAULT TREES: BFL & PFL

Having introduced our model, we can now focus our attention on property specification. In this section we will showcase some queries that one can formalize for the FT component of the FT/AT in *Figure 1*. We will do so by presenting statements in BFL, PFL and the corresponding domain specific language, LangPFL. For simplicity, we assume the original FT from [21] with the BE *Threat* replacing the AT rooted in CAT.

*BFL & PFL Properties.* Let us showcase different properties in natural language and their respective translation in BFL and PFL, starting with some BFL queries:

1) What are the MCSs for the TLE that include the presence of *Organic Matter* and deterioration of *Metallic Surface*?
$$[[MCS(WQF) \wedge OM \wedge MS]] \quad (1)$$

2) Are there MPSs for the *Disinfectant-related* subtree, given that the *DBP* and *CRD* BEs are guaranteed to fail?
$$\exists MPS(Dr)[DBP \mapsto 1, CRD \mapsto 1] \quad (2)$$

3) For all the possible configurations of BEs, are *Broken Pipes & Gaskets* plus *Loss of Pressure* sufficient for the TLE to fail?
$$\forall ((BPG \wedge LP) \Rightarrow WQF) \quad (3)$$

Note that we use the double square brackets in query (1) to signify that we want *all* the MCSs that respect the given constraints, while we use the single square brackets in query (2) to set the value of specific elements in a FT to *failed*, with 1, and to *operational*, with 0. Finally, we can ask whether a property holds for at least one/for all the possible configurations of BEs via quantifiers (∃ and ∀ respectively) as shown in queries (2) and (3). Extending BFL with PFL, we can specify some properties that include probabilities (we assume that BEs have already been assigned probability values):

1) Is the probability of TLE occurring smaller than 0.01, if the subtree rooted in *Pathway* failed?
$$Pr_{<0.01}(WQF)[Pat \mapsto 1] \quad (4)$$

2) Assume that the probability of *Organic Matter* being present equals 0.15. What would then be the probability of *Disinfectant-related* risks?
$$Pr(Dr)[OM \mapsto 0.15] \quad (5)$$

3) Assume that both *Disinfectant Loss* and *Permeation* happen with certainty. Does this imply that the probability of TLE is greater than 0.015?
$$Pr_{=1}(DL) \wedge Pr_{=1}(Pe) \Rightarrow Pr_{>0.015}(WQF) \quad (6)$$

Note that one can set arbitrary probability values for FT elements – as shown in (4) and (5) – and can specify desired thresholds for failure probabilities as per queries (4) and (6).

Furthermore, probability values for a given element can be computed anew considering what-if scenarios that account for different probabilities in the children of such an element (5). Finally, one can set assumptions on the failure probabilities of certain elements, to then check whether these values are sufficient to cause an increase exceeding given thresholds (6).

*LangPFL.* To ease usability, we showcase how these queries would be specified using the domain specific language presented in [20]. LangPFL is based on structured templates. One can specify assumptions on the status of FT elements by utilizing the *assume* keyword. These assumptions will be appropriately integrated in the translated BFL/PFL query: e.g., *set* or *setp* – to set values of FT elements – are translated with the according operators to set evidence, while other assumptions will be the antecedent of an implication. A second keyword separates specified queries from the assumptions and dictates the desired result: *compute* and *computeall* compute and return desired values, i.e., probability values and lists of MCSs/MPSs respectively, while *check* establishes if a specified property holds. Let us showcase these translations. The query in (1) would be expressed by:

$$
\begin{aligned}
&assume: \quad (7)\\
&computeall:\\
&\quad MCS[WQF] \text{ and } OM \text{ and } MS
\end{aligned}
$$

Note that the section dedicated to assumptions is empty, as we are not capturing a what-if scenario. Then *computeall* is the keyword chosen to return all MCSs with desired filters. Queries in (2) and (3) would translate to:

$$
\begin{aligned}
&assume: \quad (8)\\
&\quad \text{set DBP} = 1\\
&\quad \text{set CRD} = 1\\
&check:\\
&\quad \text{exists MPS[Dr]}
\end{aligned}
$$

$$
\begin{aligned}
&assume: \quad (9)\\
&\quad \text{set BPG} = 1\\
&\quad \text{set LP} = 1\\
&check:\\
&\quad \text{forall WQF}
\end{aligned}
$$

In (8) and (9), we see that assumptions are now populated and that we use the *check* keyword to check if the desired properties hold. Different kinds of assumptions would then be translated into different properties, as per the underlying formulations in (2) and (3). LangPFL can also handle property specification with probabilities. Queries (4), (5) and (6) would translate to:

$$
\begin{aligned}
&assume: \quad (10)\\
&\quad \text{set\_prob Pat} = 1\\
&check:\\
&\quad P[WQF] < 0.01
\end{aligned}
$$

$$
\begin{aligned}
&assume: \quad (11)\\
&\quad \text{set\_prob OM} = 0.15\\
&compute:\\
&\quad P[Dr]
\end{aligned}
$$

$$\text{assume:} \quad (12)$$
$$\text{set\_prob DL} = 1$$
$$\text{set\_prob Pe} = 1$$
$$\text{check:}$$
$$P[WQF] > 0.015$$

Operators to set evidence are now probabilistic, with *setp*, and the *compute* keyword is used to compute the probability value of the *Dr* element, given set assumptions (11). The *check* keyword remains to verify if a given property holds, as in the non-probabilistic case.

## 4 QUERYING ATTACK TREES: ATM

We now focus on the AT rooted in CAT, from *Figure 1*, by specifying some queries using ATM. Currently, this logic supports reasoning about (un)successful attacks and the formulation of properties about "cost" of attacks, "time" of an attack, both with parallel and sequential steps, "skill" needed by the attacker and "probability" of a successful attack [21]. Let us showcase some of these queries:

1) Are the costs of performing *Info Gathering and Phishing* and a *BadUSB Attack* respectively lower than 30 and at most 15?
$$Cost(IGP) < 30 \land Cost(BUA) \leq 15 \quad (13)$$

2) Is there an attack that guarantees success in executing the exploit without *Dropping USBs in the Parking Lot*?
$$\exists (EE[UPL \mapsto 0]) \quad (14)$$

3) Is it necessary that a *Privileged User Clicks on the Link* from a malicious email to mount a successful *Contamination Attack*?
$$\forall (CAT \Rightarrow UCL) \quad (15)$$

4) Are the probability of a successful *Contamination Attack* and the parallel time of attack lower than 0.010 and 30 respectively?
$$Prob(CAT) < 0.010 \land ParTime(CAT) < 30 \quad (16)$$

5) Is there an attack that ensures an attacker gains access to the system while keeping the cost under 35?
$$\exists (Cost(GA) < 35) \quad (17)$$

6) What is the minimal cost of the *Contamination Attack* assuming that the cost of the *BadUSB Attack* equals 40?
$$Cost(CAT)[BUA \mapsto 40] \quad (18)$$

As shown with these queries, ATM has a greater expressive power than BFL and PFL, as ATM can express properties that are not only concerned about probabilities but also, e.g., cost. However, constructs seen in PFL remain available. E.g., one can set arbitrary values for given metrics, as shown in (18) for "cost", can check bounds on metrics values – shown in (13), (16) and (17) – and perform quantification reasoning about some/all possible attacks. Furthermore, one can compute metric values, as in (18), or formulate queries that reason about different metrics: (16) reasons about a bounded "probability" of successful attacks, while also checking if a bound on parallel execution "time" of that attack is respected.

## 5 QUERYING FAULT TREES WITH ATTACKS

*Integrating the logics.* Having showed the capabilities of BFL, PFL and ATM we now propose a novel way to integrate these logics. The objective is to specify properties on FT/ATs that consider both the failure probabilities from the FT and how these are impacted by different metrics on the AT – e.g., success probabilities or "cost" of attacks. Writing such complex properties can be cumbersome, hence we propose to extend LangPFL to handle metrics also on ATs and FT/ATs, extending the application of *setp* operators on AT elements and introducing appropriate operators for other metrics, e.g., *setcost* for "cost". Moreover, we introduce a new construct – that we name *decorator* – employed to specify different sets of assumptions. We showcase its usage in (22). Let us introduce and comment some meaningful examples, where the part of the property related to ATM is in red. For all these properties, we assume the complete model in *Figure 1*, rooted in *WQF*.

1) Are the probabilities of TLE occurring and of an *(Un)intended Contamination* respectively lower than 0.010 and 0.005, given that the probability of a successful *BadUSB Attack* is equal to 0.12 and the probability of a privileged user clicking a malicious link is equal to 0.04?
$$Pr_{<0.010}(WQF)[CAT \mapsto \quad (19)$$
$$Prob(CAT)[BUA \mapsto 0.12, UCL \mapsto 0.04]]$$
$$\land Pr_{<0.005}(UC)[CAT \mapsto$$
$$Prob(CAT)[BUA \mapsto 0.12, UCL \mapsto 0.04]]$$

In this property, we consider the influence that the probability of success of two attack steps have on the failure probability of two FT elements. Given the what-if scenario where a *BadUSB Attack* and a privileged used clicking on a malicious email happen with probabilities 0.12 and 0.04 respectively, we can check whether the probabilities of both TLE and *(Un)intended Contamination* respect the given tresholds of 0.010 and 0.005. With our extended version of LangPFL, one would specify this query in the following way:

$$\text{assume:} \quad (20)$$
$$\text{set\_prob BUA} = 0.12$$
$$\text{set\_prob UCL} = 0.04$$
$$\text{check:}$$
$$P[WQF] < 0.010 \text{ and}$$
$$P[UC] < 0.005$$

The difference between (19) and (20) is noticeable: extending LangPFL allows practitioners to focus on property specification rather than worrying about cumbersome nesting of different logics, while still retaining needed expressivity.

In this framework, one may also consider the influence that multiple security-related what-if scenarios pose on the same FT component, e.g.:

2) Is the probability of TLE occurring lower than 0.08 in both the following scenarios: 1) when the probability of a successful *BadUSB Attack* is equal to 0.12 and the probability of a privileged user clicking a malicious link is equal to 0.04, 2) when these probabilities equal 0.34 and 0.10 respectively?

$$Pr_{<0.08}(WQF)[CAT \mapsto \qquad (21)$$
$$Prob(CAT)[BUA \mapsto 0.12, UCL \mapsto 0.04]]$$
$$\wedge Pr_{<0.08}(WQF)[CAT \mapsto$$
$$Prob(CAT)[BUA \mapsto 0.34, UCL \mapsto 0.10]]$$

This query keeps the threshold for the failure probability of TLE fixed, while varying assignments of probabilities on elements of the AT. This allows practitioners to guarantee that two different offensive scenarios will not influence the failure of TLE in undesired ways. To translate this intention in our extended version of LangPFL, we propose *decorators*: these constructs enclose a set of assumptions and allow the user to specify which properties must abide to a specific set of statements under the *assume* keyword. E.g., query (21) would be translated in the following way:

$$assume: \qquad (22)$$
$$@A1:$$
$$\quad set\_prob\ BUA = 0.12$$
$$\quad set\_prob\ UCL = 0.04$$
$$@A2:$$
$$\quad set\_prob\ BUA = 0.34$$
$$\quad set\_prob\ UCL = 0.10$$
$$check:$$
$$@A1(P[WQF] < 0.08)\ \text{and}$$
$$@A2(P[WQF] < 0.08)$$

Where *@A1* and *@A2* are two different decorators, containing two different sets of assumptions: these are declared under the *assume* keyword, as shown before. Under the *check* keyword, each part of the formula that we want to check is decorated with either *@A1* or *@A2*: assumptions in *@A1* are applied to the former occurrence of P[WQF] < 0.08, while those in *@A2* are applied to the latter.

Furthermore, we can present queries that capture the interplay between AT metrics – other than probabilities – and failure probabilities of FT elements. E.g.:

3) Is there an attack that guarantees that the failure probability of TLE would be at least 0.12 when the attacker is allowed to spend at most "cost" 30 to perform the *Contamination Attack*?

$$\exists (Cost(CAT) \leq 30\ \wedge \qquad (23)$$
$$Pr_{\geq 0.12}(WQF)[CAT \mapsto Prob(CAT)])$$

In this query, we consider both the "cost" metric and the probability of failure of TLE. The existential quantifier would range over all possible states of the leaves (both the FT and AT ones), to guarantee that there is an attack such that the cost for CAT is at most 30 and that the TLE fails with probability at least 0.12. This translates to LangPFL in the following way:

$$assume: \qquad (24)$$
$$set\_cost\ CAT \leq 30$$
$$check:$$
$$\quad exists\ P[WQF] \geq 0.12$$

In this translation, differently from (11), we see that our assumption is on cost of an AT element instead of on probability of a FT element. Finally, we can elaborate on queries (22) and (24) to construct the following:

4) Is there an attack that guarantees that: 1) the probability of TLE would be at least 0.12 and *Info Gathering & Phishing* costs at most 12 and 2) the probability of *(Un)intended Contamination* would be at least 0.08 and *Exploit and Execute* costs at most 5?

$$\exists ((Cost(IGP) \leq 12\ \wedge \qquad (25)$$
$$Pr_{\geq 0.12}(WQF)[CAT \mapsto Prob(CAT)]) \wedge$$
$$(Cost(EE) \leq 5\ \wedge$$
$$Pr_{\geq 0.08}(UC)[CAT \mapsto Prob(CAT)]))$$

This translates to LangPFL in the following way:

$$assume: \qquad (26)$$
$$@A1:$$
$$\quad set\_cost\ IGP \leq 12$$
$$@A2:$$
$$\quad set\_cost\ EE \leq 5$$
$$check:$$
$$\quad exists\ @A1(P[WQF] \geq 0.12)\ \text{and}$$
$$\quad @A2(P[UC] \geq 0.08)$$

Here, we combine the use of decorators to specify different sets of assumptions for each conjunct in the property to check (22), with the construction of queries that reason about different metrics on the AT component of FT/ATs (23).

## 6 FUTURE WORK

Our contribution opens different interesting directions for further research. Firstly, validating the framework composed by BFL, PFL and ATM on different combinations of FTs and ATs could highlight further necessities when querying models for joint safety-security analysis. Secondly, developing an implementation of the proposed approach could propel adoption of these methods in the field of reliability engineering and w.r.t. safety-critical systems. Finally, conducting hands-on tests of such an implementation with practitioners would help us in refining the framework and in tailoring it to the needs of domain experts.

### BIOGRAPHIES

*Stefano M. Nicoletti, MA*
University of Twente
Enschede, Drienerlolaan 5, 7522 NB, The Netherlands

e-mail: s.m.nicoletti@utwente.nl

Stefano M. Nicoletti is a PhD Candidate at the University of Twente, working in the ERC-funded Project CAESAR with the goal of marrying the historically separated fields of safety and (cyber)security.

*Milan Lopuhaä-Zwakenberg, Dr.*
University of Twente
Enschede, Drienerlolaan 5, 7522 NB, The Netherlands

e-mail: m.a.lopuhaa@utwente.nl

Milan Lopuhaä-Zwakenberg is an assistant professor at University of Twente (NL), studying safety and security metrics and their interplay. Before, he was a postdoc at Eindhoven University of Technology (NL) and he received his PhD from Radboud University (NL) on arithmetic geometry.

*E. Moritz Hahn, Dr.*
University of Twente
Enschede, Drienerlolaan 5, 7522 NB, The Netherlands

e-mail: e.m.hahn@utwente.nl

E. Moritz Hahn is assistant professor at the Formal Methods and Tools (FMT) group at the University of Twente within Mariëlle Stoelinga's project CAESAR. Hahn's main research interest is probabilistic model checking.

*Mariëlle Stoelinga, Prof. Dr.*
University of Twente and Radboud University
Enschede, Drienerlolaan 5, 7522 NB, The Netherlands
Nijmegen, Houtlaan 4, 6525 XZ, The Netherlands

e-mail: m.i.a.stoelinga@utwente.nl

Mariëlle Stoelinga is professor of risk management at the Radboud University and the University of Twente (NL). She is the project coordinator on PrimaVera, a large collaborative project on Predictive Maintenance in the Dutch National Science Agenda. She also received a prestigious ERC consolidator grant. She holds an MSc and a PhD degree from Radboud University and was a postdoc at the UC Santa Cruz.